\documentclass[10pt,twocolumn, nofootinbib]{revtex4-1}
\usepackage{amssymb,amsmath}
\usepackage{graphicx}
\usepackage{color}
\usepackage{hyperref}
\usepackage{listings}
\usepackage{natbib}
\usepackage{xfrac}
\usepackage{afterpage}
\usepackage{float}
\usepackage{ragged2e}
\usepackage{mathtools}
\usepackage{bm}
\usepackage{esvect}
\usepackage{upgreek}
\usepackage{lipsum}

\usepackage[T1]{fontenc}
\usepackage[utf8]{inputenc}
 
\usepackage{palatino}

\newcommand\blfootnote[1]{%
  \begingroup
  \renewcommand\thefootnote{}\footnote{#1}%
  \addtocounter{footnote}{-1}%
  \endgroup
}

\begin{document}

\title{\large \textbf{Visible blue-to-red 10 GHz frequency comb via on-chip triple-sum frequency generation}}
\author{\small{Ewelina Obrzud$^{1,2}$, Victor Brasch$^{1}$, Thibault Voumard$^{1}$, Anton Stroganov$^{3}$, Michael Geiselmann$^{3}$, Fran\c{c}ois Wildi$^{2}$, Francesco Pepe$^{2}$, Steve Lecomte$^{1}$, Tobias Herr$^{1,*}$}}
\vspace{2cm}
\affiliation{\vspace{0.25cm}\small{ \mbox{$^{1}$Swiss Center for Electronics and Microtechnology (CSEM), Rue de l'Observatoire 58, 2000 Neuch\^atel, Switzerland} \\\mbox{$^{2}$Geneva Observatory, University of Geneva, Chemin des Maillettes 51, 12901 Versoix, Switzerland} \\\mbox{$^3$LIGENTEC SA, EPFL Innovation Park, B\^atiment L, 1024 Ecublens, Switzerland}}\vspace{0.5cm}}

\begin{abstract}
\textbf{A broadband  visible blue-to-red,  10 GHz repetition  rate  frequency comb  is generated by combined spectral broadening and triple-sum frequency  generation  in  an on-chip  silicon nitride waveguide. Ultra-short pulses of 150~pJ pulse energy, generated via electro-optic modulation of a 1560~nm continuous-wave laser, are coupled to a silicon nitride waveguide giving rise to a broadband near-infrared supercontinuum. Modal phase matching inside the waveguide allows direct triple-sum frequency transfer of the near-infrared supercontinuum into the visible wavelength range covering more than 250~THz from below 400~nm to above 600~nm wavelength. This scheme directly links the mature optical telecommunication band technology to the visible wavelength band and can find application in astronomical spectrograph calibration as well as referencing of continuous-wave lasers.} 
\end{abstract}

\maketitle
\blfootnote{$^*$Tobias.Herr@csem.ch}
Optical frequency combs, coherent light sources composed of equidistant laser frequencies, provide a phase coherent link between optical and radio-frequency domains and have enabled optical frequency metrology with unprecedented precision and absolute accuracy \cite{cundiff2003}. Several precision meteorological applications such as astronomical spectrograph calibration \cite{murphy2007, steinmetz2008, li2008, braje2008, mccracken2017b} and optical clocks \cite{hinkley2013} require frequency combs in the visible wavelength range. Due to the absence of broadband laser gain media in the visible, frequency comb generation in this wavelength range is usually accomplished via nonlinear parametric frequency conversion of ultra-short pulses including frequency doubling or supercontinuum generation \cite{dudley2006}. Of particular interest are high-repetition rate systems (10-30~GHz) that could find application not only in astronomical spectrograph calibration but also in resolved comb line spectroscopy \cite{diddams2007} or optical waveform synthesis \cite{jiang2007}. Significant progress towards such systems in the visible wavelength range has been made using photonic-crystal fibres (PCF) driven by mode-filtered frequency-doubled fibre lasers \cite{probst2016} and titanium-sapphire lasers \cite{glenday2015, mccracken2017a} as well as using frequency doubled electro-optic modulation combs \cite{metcalf2019}. Despite these efforts covering the visible wavelength range, in particular the short wavelength portion towards 400~nm and below, remains challenging.

Recently, integrated chip-based waveguides have emerged as a nonlinear optical platform that offers unique opportunities both for spectral broadening and harmonic generation. These waveguides are lithographically defined and can be fabricated from various materials, including silicon \cite{Leuthold2010}, silicon nitride \cite{moss2013}, aluminum nitride \cite{jungtang2016} and lithium niobate \cite{zhang2017}. Their high material nonlinearity ($\upchi ^{(2)}\, \mathrm{and/or}\, \upchi ^{(3)}$) and small mode cross-section result in highly efficient nonlinear optical frequency conversion. In addition, waveguides offer a flexible dispersion design that can be tailored to a given application by adapting their geometry. Recent progress in broadband frequency conversion in waveguides has given rise to broadband spectra in the ultra-violet and visible wavelength ranges. Specific methods used include $\upchi^{(3)}$-based supercontinuum generation \cite{mayer2015, porcel2017b, oh2017, okawachi2018}, combined $\upchi^{(3)}$-$\upchi^{(2)}$-based supercontinuum and sum-frequency generation in waveguides with intrinsic \cite{langrock2007, phillips2011, iwakuni2016, hickstein2017, carlson2017, yoshii2019, liu2019, vasilyev2019, chen2019, yu2019} or optically induced second order nonlinearity \cite{hickstein2019}. Covering the 400--600~nm wavelength range remains however exceedingly challenging, in particular for multi-GHz pulse repetition rates.

\begin{figure}[!ht]
	\centering
	\includegraphics[width=0.48\textwidth]{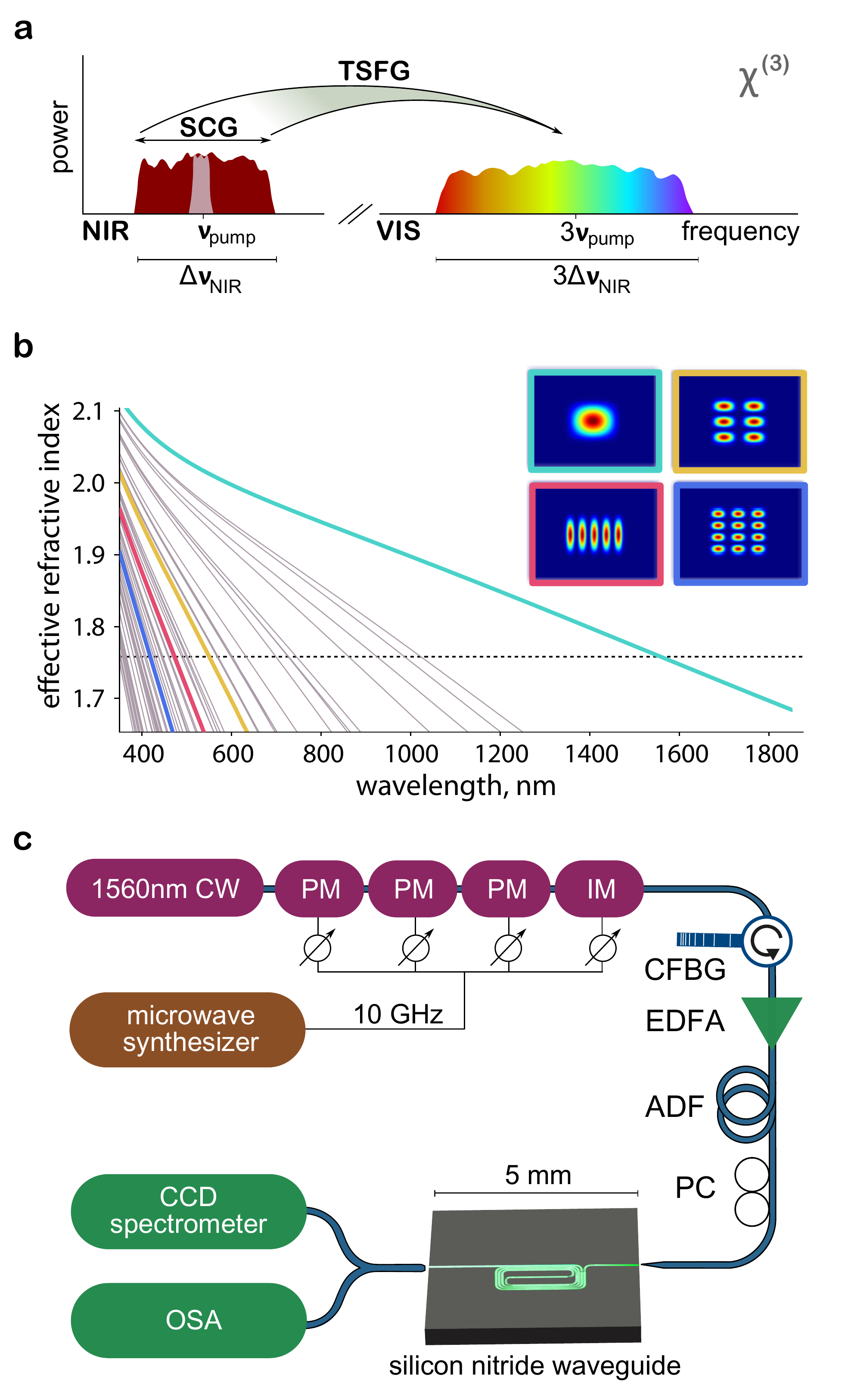}
	\caption{\small(a) Conversion from near-infrared (NIR) to visible (VIS) optical frequencies $\nu$ in a $\chi ^{(3)}$-nonlinear waveguide pumped with ultra-short pulses. SCG - supercontinuum generation, TSFG - triple-sum frequency generation. (b) Effective refractive indices $n_\mathrm{eff}$ for all waveguide modes of one polarisation found via finite-element simulation. The dashed line corresponds to $n_\mathrm{eff}$ at the pump wavelength of 1560~nm. The insets show the mode profiles for selected modes as indicated by the colour. (c) Experimental setup comprising an electro-optic pulse generator and a chip-based silicon nitride waveguides. The generated spectra are analysed with a CCD spectrometer and optical spectrum analysers (OSAs). PM - phase modulator, IM - intensity modulator, CFBG - chirped fibre Bragg grating, EDFA - erbium-doped fibre amplifier, ADF - sign-alternating dispersion fibre, PC - polarisation controller.}
	\label{setup}
\end{figure}

Here, we explore near-infrared supercontinuum generation and direct broadband frequency conversion to the visible of a 10 GHz repetition rate frequency comb in a silicon nitride waveguide relying only on $\chi ^{(3)}$-nonlinear processes (Fig. \ref{setup}a). First, a broadband supercontinuum is generated around the near-infrared pump frequency. Second, triple-sum frequency generation (TSFG), i.e. summing of three optical frequencies, results in a visible frequency comb with the same comb line spacing and approximately three times the width of the near-infrared spectrum. With this method we generate a broadband visible spectrum spanning more than 250 THz from below 400~nm to above 600~nm with on average 0.4~nW of power per mode.

 \begin{figure}[!ht]
	\centering
	\includegraphics[width=0.48\textwidth]{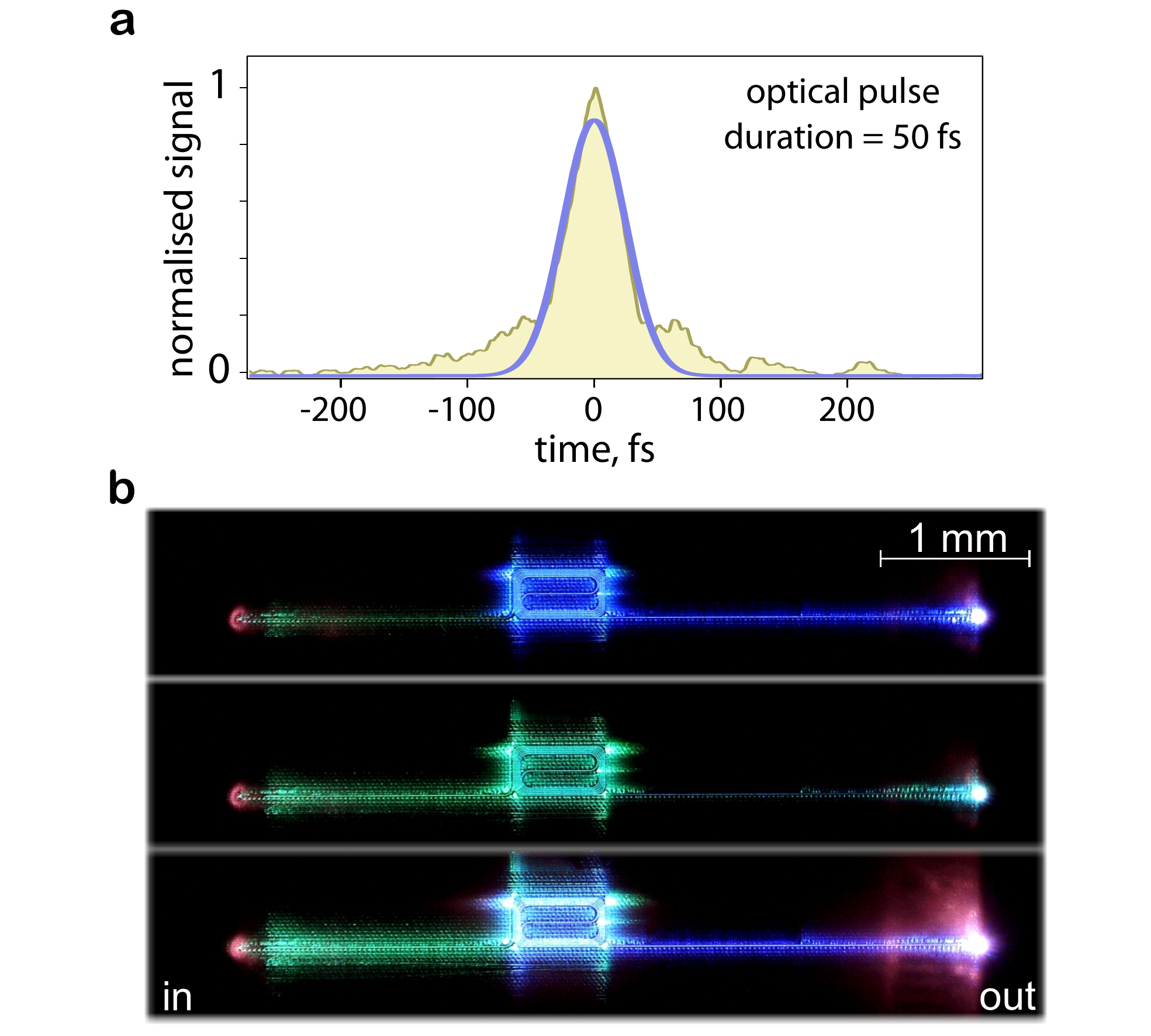}
	\caption{\small (a) Autocorrelation trace of the optical pulses coupled into the waveguide (yellow), indicating a pulse duration of 50~fs assuming a Gaussian pulse shape (blue fit). (b) Photographs of the waveguide during operation with different input polarisation.}
	\label{autocorrelation}
\end{figure}

\begin{figure*}[!ht]
	\centering
	\includegraphics[width=1\textwidth]{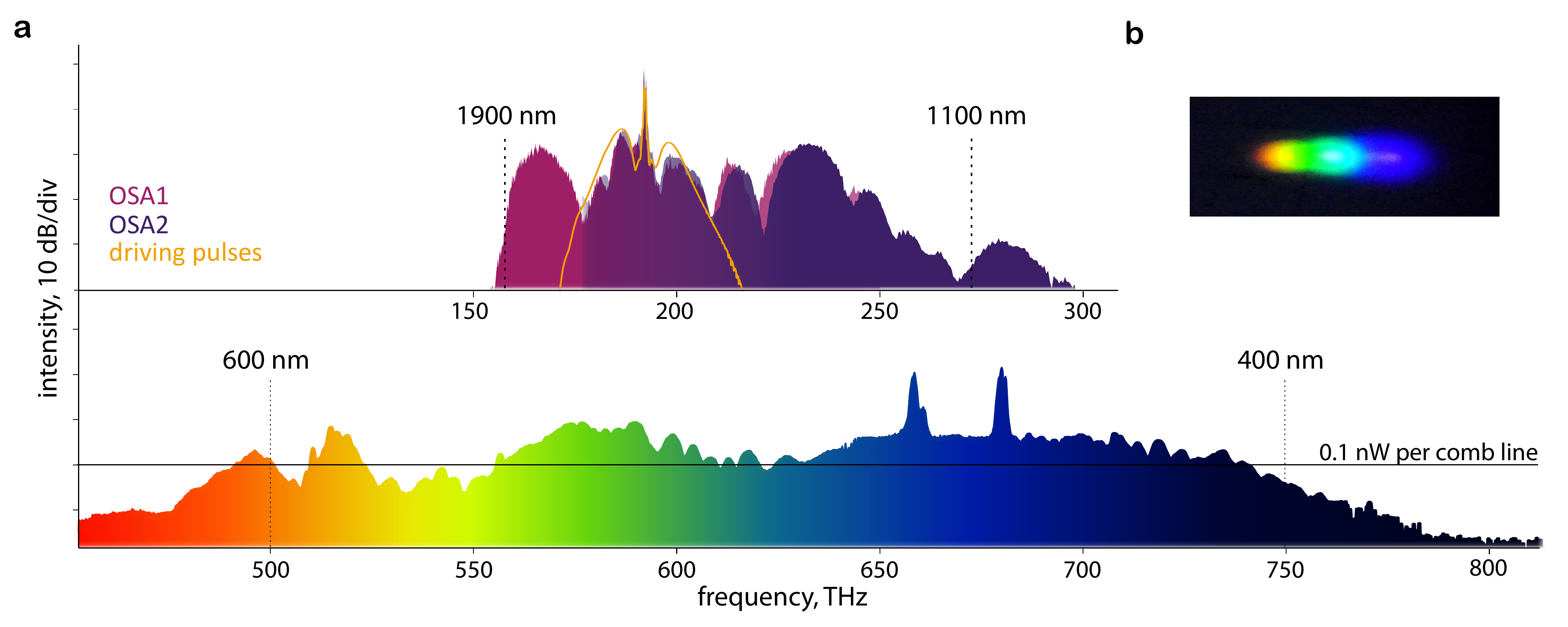}
	\caption{\small (a) Top: Near-infrared supercontinuum spectrum (pink and violet traces) generated by 10~GHz repetition rate, 50~fs duration and 150~pJ energy pulse, recorded with two optical spectrum analysers (OSAs). The pump spectrum is indicated by the orange trace. Bottom: Visible spectrum recorded after the silicon nitride waveguide with a CCD spectrometer. The frequency axes in both plots are on the same scale. b) Image of the visible light after the silicon nitride waveguide dispersed by a diffraction grating. The horizontal line indicates the power level of 0.1~nm per comb mode. Dashed vertical lines mark specific wavelengths.}
	\label{spectrum}
\end{figure*}

\begin{figure}[!ht]
	\centering
	\includegraphics[width=0.48\textwidth]{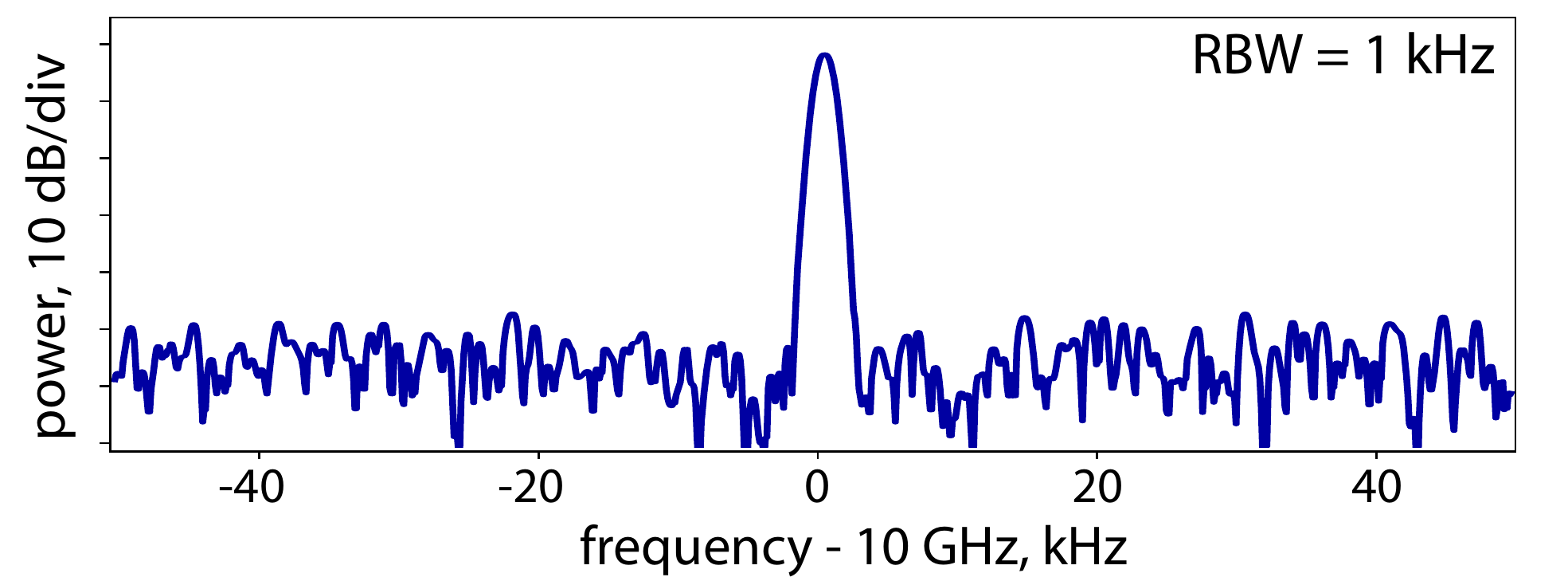}
	\caption{\small Beatnote measured for the generated spectrum in the wavelength range from 400~nm to 450~nm.}
	\label{beatnote}
\end{figure}

Nonlinear optical frequency conversion in parametric processes requires the phase-matching condition to be satisfied, i.e. sums of all wave vectors involved in the nonlinear mixing process $\sum_{\mathrm{i}} \vv{\mathbf{k}}_\mathrm{i}=0$ where $|\vv{\mathbf{k}}_\mathrm{i}| = \mathrm{n_{eff} \cdot \upomega _i/c}$ is the wave vector of the mixing frequencies with $\mathrm{n_{eff}}$ the effective index of refraction  at $\mathrm{\upomega _i}$ (including both the material and geometric dispersion). Various schemes are used to provide phase-matching including birefringent phase-matching, quasi-phase matching and modal phase-matching \cite{rao2004}, which relies on modal overlap and phase matching between fundamental and higher order modes. Additionally, when the pump source is a train of ultra-short pulses, matching the group velocity dispersion (GVD) of the fundamental and harmonic waves, which can be achieved by adapting the geometry of a waveguide \cite{foster2006}, can further increase the efficiency.

In our experiment, we use an integrated low pressure chemical vapor deposition (LPCVD) silicon nitride ($\mathrm{Si_3N_4}$) waveguide offering a high Kerr-nonlinearity of $\mathrm{2.4 \cdot 10^{-19}\, m^2/W}$ \cite{ikeda2008} with low propagation losses. The large material band gap of 5~eV and transparency window make it an ideal platform for spectral transfer into the visible. While the waveguide does not allow for GVD-matching between the near-infrared and the short-wave visible wavelength ranges, the highly multi-mode structure of the waveguide at visible wavelengths permits modal phase-matching as illustrated in Figure~\ref{setup}b, where for each wavelength a mode with an effective refractive index close to the one at the pump wavelength (1560~nm) is found (dashed line). 

As a driving pulse source we use an electro-optic high-repetition rate ultra-short pulse generator operating in the near-infrared erbium gain window \cite{kobayashi1988, Torres-Company2014, beha2017, okubo2018, obrzud2018, nakamura2019, kashiwagi2019}. Our pulse source \cite{obrzud2018}, illustrated in Figure \ref{setup}c, offers great flexibility in the repetition rate choice (5-15 GHz) and relies on mature polarisation maintaining off-the-shelf optical telecommunication components assuring robust operation without the necessity of any alignment: A narrow-linewidth continuous-wave (CW) laser at 1560~nm is phase- and intensity-modulated by three phase modulators and one intensity modulator, respectively, driven by an external microwave synthesizer at a repetition rate of 10~GHz. The phase of the RF signal at each modulator is adjusted such that the phase modulation builds up coherently and the intensity modulation carves out light with the correct sign of chirp. By compensating the chirp via a chirped fibre-Bragg grating (CFBG) ($\approx 2$~ps/nm), a train of pulses with a duration of approximately 2~ps is formed. The generated initial flat-top spectrum spans approximately 6~nm and has an average power of a few mW. The light is then amplified in an erbium-doped fibre amplifier (EDFA) up to approximately 4.5~W of average power and the pulse width compresses down to 230~fs. In order to further decrease the pulse duration, and so ensure a high pulse peak power for efficient frequency conversion, the light is sent through a combination of nonlinear optical fibres with alternating GVD sign. Optical pulses are first injected into a positive GVD fibre where they broaden spectrally via SPM while acquiring a linear chirp. The following negative GVD part is adjusted in length so that the pulses recompress by the effect of dispersion. The output pulses reach a full-width at half-maximum (FWHM) duration of 50~fs assuming a Gaussian pulse shape (Fig.\ref{autocorrelation}a). To optimise the polarisation, a polarisation controller is included before the waveguide.

The optical pulses are then coupled into a silicon nitride waveguide using a lensed fibre mounted on a 3-axis translation stage for fibre-waveguide alignment. The 14~mm-long waveguide has the width and height of 1000~nm and 800~nm, respectively, which are chosen to provide close to zero GVD at 1560~nm for efficient broadband near-infrared supercontinuum generation. The propagation losses are approximately 0.2~dB/cm. The outcoupled light is collected by a 100~$\upmu$m core multi-mode fibre with a transmission loss below 1~dB in the 400~nm to 2~$\upmu$m wavelength range. The visible and near-infrared spectra are recorded by a CCD spectrometer and two optical spectrum analyzers, respectively. Based on the fibre-to-waveguide coupling efficiency we estimate that pulses with an energy of approximately 150~pJ are coupled into the silicon nitride waveguide.

Figure \ref{autocorrelation}b shows images of the waveguide during operation for different settings of the input polarisation. Changing the polarisation of the injected pulses results in different mode excitation in the waveguide and influences the generated spectrum. A narrowband green triple-sum spectrum of the input pulses is instantaneously generated in the waveguide whereas broadband visible and near infrared spectra only emerge after a longer propagation distance inside the waveguide. The resulting spectra are shown in Fig. \ref{spectrum}a. The top panel shows the driving pulses (orange trace) and the near-infrared spectrum that was recorded with two optical spectral analysers. It spans approximately 100~THz, from 1200~nm to 1900~nm. The bottom panel represents the visible spectrum recorded with a CCD spectrometer. It extends across more than 250 THz, from below 400 to above 600 nm covering the short wavelength visible wavelength domain. The horizontal dashed line in Figure~\ref{spectrum} indicates the power level of 0.1~nW per mode, which is sufficient for CW laser referencing and astronomical spectrograph calibration. The two high intensity spectral features in the blue wavelength domain are likely associated with a particularly strong mode-overlap between the phase matched fundamental and higher order modes in the near-infrared and visible range, respectively. Figure \ref{spectrum}b shows a photograph of the visible light at the output of the collecting fibre diffracted by a grating. The length of the waveguide was optimised for the broadest spectral width and the maximum power output in the visible wavelength range. Shorter waveguides resulted in narrower spectral bandwidth in both the near-infrared and visible spectral band. On the other hand, longer waveguides produced visible spectra with lower output powers likely caused by additional losses due to scattering and waveguide bending. 
Because the resolution of the CCD spectrometer does not allow resolving the comb lines in the visible, we measure the beatnote in the blue part of the spectrum after sending it through a 400--450~nm bandpass transmission filter (Fig. \ref{beatnote}). The observed beatnote with a signal-to-noise of 50 dB (1 kHz resolution bandwidth) confirms the comb nature of the visible spectrum.

To summarize, we have demonstrated simultaneous generation of broadband, high-repetition rate near-infrared and visible wavelength frequency combs via combined supercontinuum and triple-sum-frequency generation in a silicon nitride waveguide. This scheme provides direct access to the visible blue-to-red wavelength domain from the technologically mature optical telecommunication band. The achieved power levels per mode are sufficient for laser referencing in atomic and molecular physics as well as astronomical spectrometer calibration, where the demonstrated results can help to overcome current challenges. We anticipate that reduced waveguide bending and scattering losses, optimised visible wavelength output couplers, resonant approaches \cite{herr2018} and advanced waveguide dispersion engineering \cite{foster2006, Guo2019} can further increase the conversion efficiency for applications where this is required.

\paragraph*{\textbf{Funding}}
This work was supported by the Swiss National Science Foundation (grants 200020\textunderscore 182598, 200020\textunderscore 166227 and 200020\textunderscore 184618), the Swiss National Science Foundation and Innosuisse (20B2-1\_176563), the Technology Platform of the National Centre for Competence in Research “PlanetS”, the Swiss Space Office and the Canton of Neuch\^atel.


\end{document}